\documentclass[aps,pra,twocolumn,amsmath,showpacs,letterpaper,floatfix]{revtex4}
\usepackage{graphicx}           % Include eps- und ps-Graphics
\newcommand{\ket}[1]{\left| #1\right\rangle}

\newcommand{\ketbra}[3][]{\left|#2\right\rangle_{#1}\!\left\langle#3\right|}

\newcommand{\iea}[0]{{\it et al.~}}

\begin{document}

\title{A bridge between the single-photon and squeezed-vacuum state}

%\author{N Jain$^1$, S Huisman$^1$\footnote{Present address: Complex Photonic Systems (COPS), Faculty of Science and Technology and MESA+ Institute for Nanotechnology, University of Twente, The Netherlands}, E Bimbard$^1$\footnote{Permanent address: Department of Physics, Ecole Normale Sup\'erieure in Paris, France} and A I Lvovsky$^1$}

\author{Nitin Jain\footnote{Present address:
Max Planck Institute for the Science of Light, G\"unther-Scharowsky-Str. 1, 91058 Erlangen, Germany}, S. R. Huisman\footnote{Permanent address: Complex Photonic Systems (COPS), Faculty of Science and Technology and MESA+ Institute for Nanotechnology, University of Twente, 7500 AE Enschede, The Netherlands}, Erwan Bimbard\footnote{Present address: Department of Physics, Ecole Normale Sup\'erieure in Paris, France} and A. I. Lvovsky\footnote{Email: LVOV@ucalgary.ca}}

\address{{Institute for Quantum Information Science, University of Calgary, Calgary, Alberta T2N 1N4, Canada}}
%\ead{lvov@ucalgary.ca}
\date{\today}

\begin{abstract}
The two modes of the Einstein-Podolsky-Rosen quadrature entangled state generated by parametric down-conversion interfere on a beam splitter of variable splitting ratio. Detection of a photon in one of the beam splitter output channels heralds preparation of a signal state in the other, which is characterized using homodyne tomography. By controlling the beam splitting ratio, the signal state can be chosen anywhere between the single-photon and squeezed state.
\end{abstract}


%\submitto{njp}
\maketitle

\paragraph{Introduction} \label{sec:Introduction}
Nonclassical states of light form an important tool in testing fundamental quantum physics and quantum information processing. Among the most basic quantum optical states are the squeezed-vacuum states, that exhibit a reduced noise in amplitude or phase, and the photon number states (Fock states) with a well-defined energy. The former were among the first nonclassical states of light to be generated experimentally and are the basis for a variety of continuous-variable quantum information protocols \cite{cv-book}. Such states feature positive Gaussian Wigner functions and the field quadrature noise is strongly phase dependent. Fock states, especially the single-photon state are natural candidates for encoding quantum information in the discrete-variable setting (qubits) \cite{kok}. Their Wigner functions show non-Gaussian characterisitics such as oscillations and negativities and are phase-independent \cite{spf,spf-bellini,FockNewAge}.

These ``discrete-variable'' and ``continuous-variable'' pillars of quantum-optical technology developed separately for a long time, each associated with its own set of production and detection methods as well as applications. Recently, these pillars have been bridged by a variety of experiments extending the range of the Hilbert space accessible to quantum technology beyond the framework of either domain  \cite{photonadded,schcat06,walmsley09,CVdistill,quantech2}.

In this paper we report an experiment that demonstrates the connection between the discrete and continuous domains of quantum optics very explicitly. We start with a two-mode squeezed state produced through parametric down-conversion in a spectrally and spatially degenerate, but polarization-nondegenerate configuration. The two modes of this state are overlapped on a beam splitter of variable splitting ratio, formed by a half-wave plate (HWP) and a polarizing beam splitter (Fig.~1). One of the output channels of the beam splitter (\emph{trigger}) is subjected to a measurement by a single-photon detector. Conditioned on a photon-detection event, and dependent on the reflectivity of the beam splitter, a particular quantum state of light in the other channel (\emph{signal}) is prepared.

%a field quadrature measurement of theperformed by means of a homodyne detector. This procedure is repeated %multiple times so that the signal state can be completely characterized using homodyne %tomography \cite{alex-rvw}.

If the beam splitter reflectivity $R$ equals 0 or 1, the two-mode squeezed state remains unaltered. If the degree of squeezing is small (as is the case in our experiment), a ``click" of the single-photon detector heralds the preparation of the single photon in the signal channel with a high probability. On the other hand, a beam splitter with a reflectivity of $1/2$ converts the two-mode squeezed state into a tensor product of two squeezed states. Independent of the measurement result in the trigger channel, the signal state is squeezed vacuum. In this way, by simply rotating the waveplate that determines the beam splitter reflectivity, we can choose the output to be the squeezed-vacuum or single-photon state, or anything in between.

\paragraph{Experiment}
\begin{figure}
\centering
\includegraphics[width=\columnwidth]{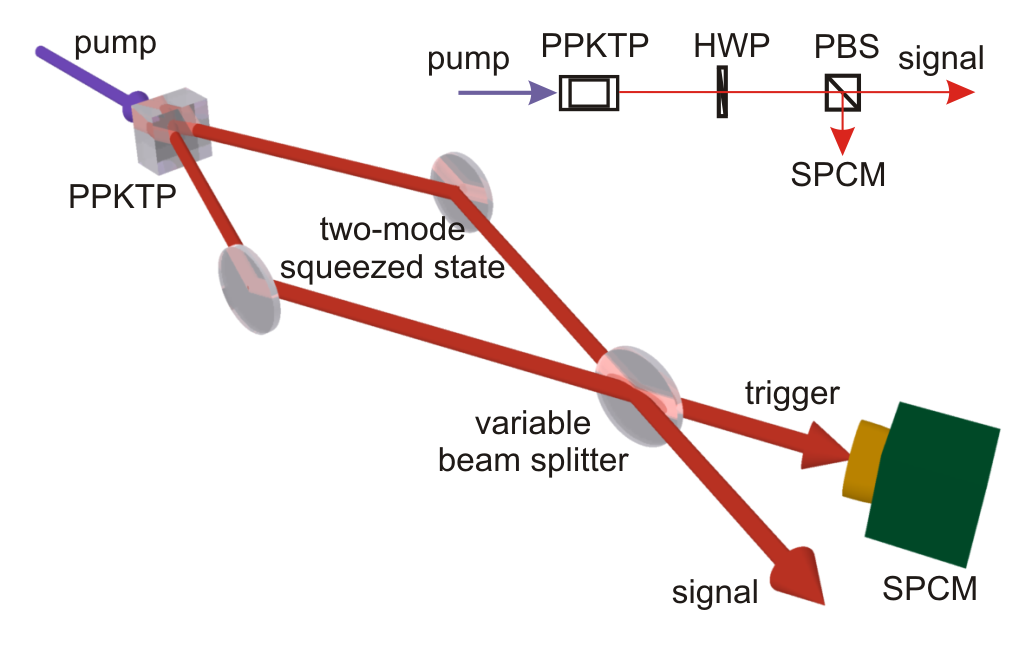}
\caption{Basic schematic of the experimental setup. The figure is for illustration purpose only; the inset shows the actual implementation of parametric down-conversion and the variable beam splitter. HWP, half-wave plate; PBS, polarizing beam splitter; SPCM, single-photon counting module. The down-conversion is spatially and spectrally-degenerate, but polarization-nondegenerate.}\label{setup}
\end{figure}
The operation of the setup is explained in detail in Ref.~\cite{FockNewAge}; a brief description relevant to this article follows. A mode-locked Ti:Sapphire laser (Coherent MIRA 900) emits transform-limited pulses of width around 1.7 ps and a repetition rate of about 76 MHz. The laser output, centered at a wavelength $\sim$790 nm, is frequency doubled and focussed onto a periodically-poled potassium titanyl phosphate (PPKTP) crystal, which is phase-matched for a type II collinear down conversion. The two collinear output channels are mixed on the variable beam splitter, producing the trigger and signal modes. The trigger mode is subjected to spectral selection using a 0.3-nm bandwidth interference filter and spatial filtering with a single-mode optical fiber followed by measurement with a single photon counting module (Perkin-Elmer SPCM-AQR-14-FC).

The state in the signal mode, heralded by the detection event in the trigger channel, is characterized using optical homodyne tomography\cite{alex-rvw}. The local oscillator for homodyne detection is obtained from the master laser output. Aichele \iea~\cite{aichele02} describe in detail the method for matching the spatio-temporal mode of the local oscillator pulses to the signal state. The difference photocurrent signal from the homodyne detector is amplified and digitized by an acquisition card (Agilent Acqiris DP211) that is triggered by the output of the SPCM. For each position of the HWP, a set containing $10^6$ quadrature samples of the signal state plus $8\times10^6$ samples from 8 neighbouring pulses, is acquired. The signal channel is then blocked and a set containing $9\times10^6$ uniform vacuum state samples is acquired in order to calibrate the quadrature scale.

\begin{figure}
\centering
\includegraphics[width=\columnwidth]{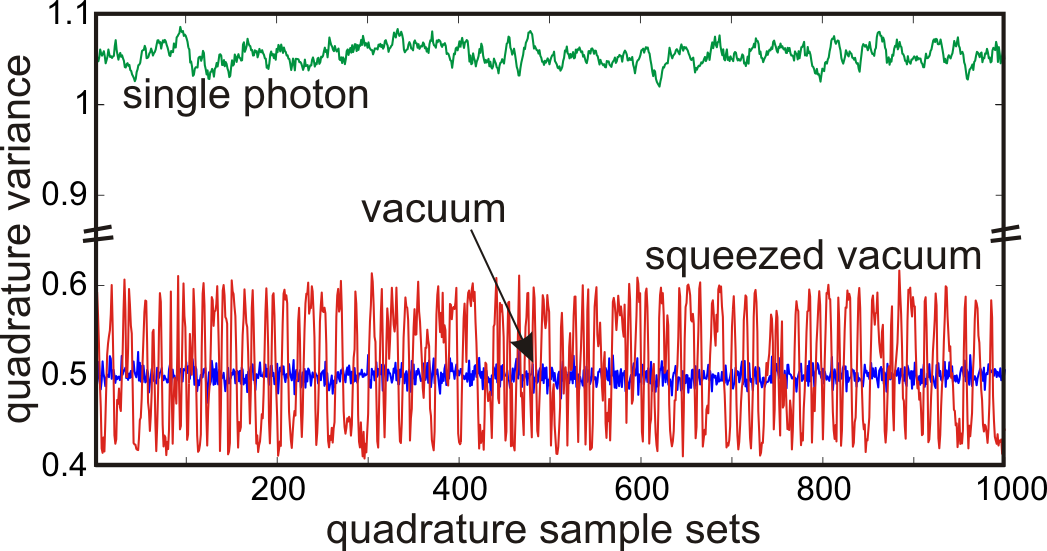}
\caption{Quadrature noise of the experimentally observed vacuum, squeezed-vacuum and single-photon Fock states. The plot displays the variances of 1000 sets, each containing 1000 quadrature samples acquired while varying the local oscillator phase. This phase is noticeably affected by mechanical instabilities and air movement. The observed quadrature noise is affected by preparation and detection inefficiencies.}\label{var}
\end{figure}

\begin{figure}
\centering
\includegraphics[width=\columnwidth]{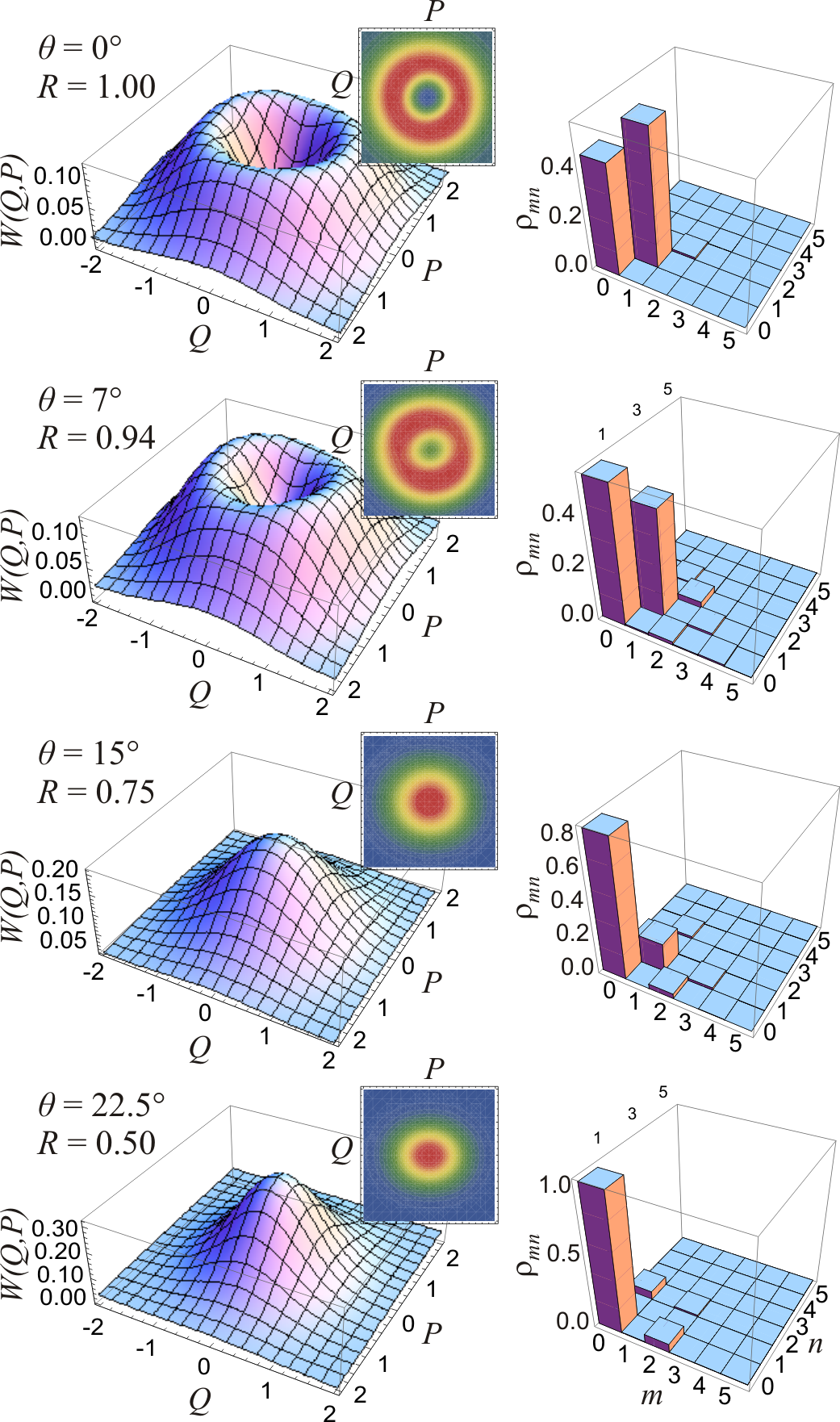}
\caption{Experimentally reconstructed Wigner functions (left column) and density matrices (absolute values, right column). The angle $\theta$ of the half-wave plate and the corresponding beam splitter reflectivity (given by $R=\cos ^2 2\theta$) are indicated for each set. The insets show the contour diagrams associated to that specific Wigner function.}\label{dm}
\end{figure}

Figure \ref{var} shows the experimentally measured quadrature variance $\Delta^2 Q_{\phi}$ for the single-photon state and the squeezed-vacuum state, acquired at the HWP angles of $0^{\circ}$ and $22.5^{\circ}$, respectively. Each point is calculated from a set of 1000 quadrature samples, acquired within a time interval of $\sim 0.02$ s, during which the local oscillator phase does not change significantly. The quadrature noise of a pure Fock state $\ket{n}$ is independent of the local oscillator phase $\phi$ and given by $\Delta^2 Q = (2n+1)/2$. The data acquired in our experiment corresponds, with a high precision, to the state $\hat\rho=\eta_{\ket 1}\ketbra 11+(1-\eta_{\ket 1})\ketbra 00$, where $\eta_{\ket 1}$ is the overall efficiency \cite{spf} taking into account the linear losses, dark counts of the single-photon detector, mode mismatch with the local oscillator \cite{aichele02} and electronic noise of the homodyne detector \cite{Dallas}. We observe $\Delta^2 Q=1.05$, which implies $\eta_{\ket 1}=0.55$. Small fluctuations of the variance visible in Fig.~\ref{var} are of statistical nature and not correlated with the local oscillator phase.

Conversely, an ideal squeezed-vacuum state is phase-sensitive, and its variance as a function of the phase can be expressed as $\Delta^2 Q_{\phi}  = A+B\cos 2\phi$. If the experiment were perfectly efficient, the minimum uncertainty condition $(A+B)(A-B)=1/4$ would hold. In practice, because of a nonunitary efficiency, this product is larger.

\begin{figure}
\centering
\includegraphics[width=\columnwidth]{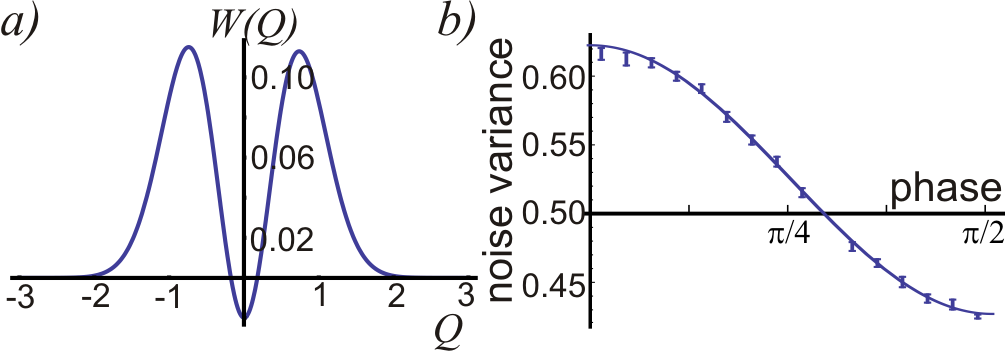}
\caption{The cross-section of the Wigner Function of the single-photon state (a) and the quadrature noise of the squeezed state as a function of the phase (b), obtained from their respective experimentally reconstructed density matrices. The squeezing variance features a solid curve obtained from maximum-likelihood reconstruction, while the points with error-bars from 0 to $\pi$/2 are representative of the binned raw quadrature data. The error bars correspond to $\sigma_i\sqrt{2/N_i}$, $\sigma_i$ being the width of a Gaussian distribution from $N_i$ samples in each bin \cite{bookcompmthd}.
}\label{sqzvarfockcrs}
\end{figure}

In fact, the above expression for the quadrature noise variance holds for all data sets at any HWP angle (in the extreme case of $0^\circ$ angle, $B=0$). We utilize this dependence to determine the local oscillator phase for each set of 1000 quadrature samples. In this way, for each position of the wave plate, we obtain $10^6$ quadrature-phase pairs associated with the signal state. We use then reconstruct the state using a quantum likelihood-maximization algorithm \cite{Lvovsky2004}, without correcting for detection inefficiency.

Figure \ref{dm} presents the density matrices and corresponding Wigner functions reconstructed from the experimentally retrieved data. One can observe a gradual transition from the single-photon to the squeezed-vacuum state. For the $R=0.5$ case, the observed density matrix approximates that of a superposition of the vacuum and two-photon states, and the corresponding Wigner function exhibits squeezing of the momentum quadrature. On the other hand, for $R=1$ the only significant density matrix elements are the diagonal elements corresponding to the single-photon and vacuum states.

%\begin{figure}
%\centering
%\includegraphics[width=\columnwidth]{TPS.png}
%\caption{Biphoton spectrum, calculated from the parameters of the PPKTP crystal and the pump pulse. }\label{tps}
%\end{figure}

These extreme cases are further elaborated in Fig.~\ref{sqzvarfockcrs}. The left pane shows a cross-section of the Wigner function of the single-photon state featuring negative values around the phase-space origin that are characteristic of this state. In the right pane, the variance of the quadrature noise as a function of the phase is displayed for the squeezed vacuum, showing a reduction of 0.62 dB below the standard quantum limit.

The efficiency of the squeezed-vacuum measurement can be evaluated from the deviation of the observed noise variances from the minimum uncertainty limit \cite{pulsqz}:
\begin{equation}\label{etacmsq}
\eta_{sq} = \frac{2\langle Q^2_+ \rangle + 2\langle Q^2_- \rangle - 4\langle Q^2_+ \rangle \langle Q^2_- \rangle - 1}{2\langle Q^2_+ \rangle + 2\langle Q^2_- \rangle - 2}
\end{equation}
where $\langle Q^2_+ \rangle=A+B$ and  $\langle Q^2_- \rangle=A-B$ are, respectively, the highest and the lowest quadrature variances. Our data yields $\eta_{sq} = 29$\%. The observed discrepancy between the efficiencies of the single-photon and squeezed-vacuum states can be accounted to the multimode character of pulsed squeezed vacuum associated with the correlated character of the biphoton spectrum \cite{pulsqz,alexindepsqz}. The PPKTP crystal used in our experiment (2 mm length, $\sim$ 9 $\mu$m poling period) produces a spectrum which can be expected to lead to a 94\% inefficiency in addition to all the factors that contribute to $\eta_{\ket 1}$. This number cannot fully account for the difference between $\eta_{\ket 1}$ and $\eta_{sq}$. Additional losses are likely to be associated with the spatial multimode character of the biphoton.

\paragraph{Summary}
We have reported an experiment bridging the discrete and continuous domains of quantum optics. Employing conditional detection and homodyne tomography, various quantum states ranging from the single-photon Fock state to the squeezed-vacuum state can be prepared and characterized. The overall efficiency for tomographic reconstruction of the squeezed-vacuum state is below that of the single-photon state, likely due to the spatially and spectrally multimode character of the biphoton generated in parametric down-conversion.

%\section{Acknowledgements}
This work was supported by NSERC, CIAR, AIF, CFI and Quantum\emph{Works}. We thank M. Lobino for helpful discussions.

%\section*{References}
%\verb"\begin{thebibliography}{24}"

\end{document}